\documentclass[12pt, preprint]{aastex}
\usepackage{epsf}
\usepackage{epsfig}
\usepackage{times}
\newcommand{\etal}{{et al}\/.}

\begin{document}
\slugcomment{Draft of \today}
\shorttitle{Shock heating in NGC\,3801}
\shortauthors{J.H. Croston \etal}
\title{Shock heating in the nearby radio galaxy NGC\,3801}
\author{J.H. Croston\altaffilmark{1}, R.P. Kraft\altaffilmark{2}, M.J. Hardcastle\altaffilmark{1}}
\altaffiltext{1}{School of Physics, Astronomy and Mathematics, University of
Hertfordshire, College Lane, Hatfield AL10 9AB, UK}
\altaffiltext{2}{Harvard-Smithsonian Center for Astrophysics, 60 Garden Street, Cambridge, MA~02138, USA}

\begin{abstract}
 We report the {\it Chandra} detection of shock-heated shells of hot
 gas surrounding the radio lobes of the nearby ($D_{L} \sim 53$ Mpc)
 low-power radio galaxy NGC\,3801. The shells have temperatures of 1
 keV and 0.7 keV, compared to an ISM temperature of 0.23 keV. The
 estimated expansion speed of the shells is $\sim 850$ km s$^{-1}$,
 corresponding to a Mach number of $\sim 4$. This is the second X-ray
 detection of strong shocks produced by a low-power radio galaxy, and
 allows us to measure directly the contribution of shock heating to
 the radio galaxy's total energetic input to the ISM. We show that the
 gas properties of the shells and surrounding ISM are consistent with
 the Rankine-Hugoniot shock jump conditions. We estimate the energy
 stored in the hot gas shells (thermal + kinetic energy) to be $1.7
 \times 10^{56}$ ergs, which is equivalent to the thermal energy of
 the ISM within $\sim 11$ kpc of the galaxy centre, and a factor of
 $\sim 25$ larger than the inferred $P$d$V$ work required to inflate
 the lobe cavities, indicating that energy transfer from the AGN to
 its environment is dominated by shock heating during this stage of
 radio-source evolution. Our results provide direct evidence that
 shock heating in the early supersonic phase of FRI radio-source
 expansion can have important long-term effects on the properties of
 the host galaxy ISM. Finally, we discuss the merger history of
 NGC\,3801, the fuelling of its AGN and the role of this type of
 system in feedback models.

\end{abstract}
\keywords{galaxies: active -- X-rays: galaxies -- X-rays: quasars
-- radiation mechanisms: non-thermal}

\maketitle

\section{Introduction}
\label{intro}

Shock heating by radio galaxies is thought to be an important means of
transferring energy from active galactic nuclei to their environments,
but direct evidence for this process has been difficult to find. The
first example of a shock associated with expanding radio-galaxy lobes
was found in the nearest low-power (FRI) radio galaxy, Centaurus A
(Kraft et al. 2003), and this remains the only case to date in which a
high Mach number shock can be seen in direct contact with a radio
lobe. Other examples of shocks have been found in clusters, where they
are often detached from the radio lobes and always considerably weaker
(e.g. M87: Forman et al. 2005, Cygnus A: Wilson et al. 2006; Hydra A:
Nulsen et al. 2006); these results confirm that energy input via
shocks is an important mechanism for AGN feedback on scales ranging
from individual galaxies to galaxy clusters. It is therefore essential
to be able to constrain the physical properties of radio-lobe shocks,
and to understand in what circumstances they occur, so as to be able
to measure the total energy contribution from radio galaxies to their
environments.

The detection of shocked gas in the low-power (FRI) radio galaxy
Centaurus A (Kraft et al. 2003), suggests that the early stages of
radio-source expansion will have a dramatic effect on the host galaxy
ISM. Well-studied, large FRI sources are not thought to be
overpressured \citep[e.g.][]{c03}; however, it is likely that all
radio sources go through an early overpressured phase of supersonic
expansion, producing shocks, before achieving pressure balance
\citep[e.g][]{hei98}. While the highly supersonic phase of expansion
must last for only a small fraction of the source's lifetime, the
impact on the surrounding gas properties of the shock heating it
produces may be dramatic, since the energy input from such a source is
a significant fraction of the gravitational energy of the gas: in the
case of Cen A, the shock-heated gas has a thermal energy of $4.2
\times 10^{55}$ ergs, which is a few tens of percent of that of the
ISM (and its kinetic energy is significantly higher -- see
\citet{kra03}). Low-power FRIs similar to Cen A are expected to be
relatively common, and so the shock detection in Cen A leads to a
prediction that many elliptical galaxies experience the effects of
shock heating by supersonically expanding radio lobes. It is therefore
essential to establish whether the shock conditions observed in Cen A,
which do not agree with standard predictions, are likely to be typical
of the effects of young FRI sources on their environments.

In this paper we report the {\it Chandra} detection of shocked gas
shells surrounding the radio lobes of NGC\,3801: this is the second
such detection in a nearby radio galaxy. NGC\,3801 is a nearby ($z =
0.0113$), isolated E/S0 galaxy, which hosts a small double-lobed radio
source of similar morphology to Cen A. Optically, NGC\,3801 is
disturbed, with a prominent dust lane and secondary dust features, as
shown in HST WFPC2 images \citep{ver99}. Recent millimeter-wave
observations with BIMA \citep{das05} have revealed the presence of a
central rotating molecular gas ring or disk, as well as a gas clump
likely to be infalling merger-related material that may be interacting
with the radio jet. These results indicate that NGC\,3801 has recently
undergone a merger; our {\it Chandra} data therefore also provide an
opportunity to investigate the relationship between merger activity,
AGN fuelling and feedback via radio outbursts. Since small FRI sources
like Cen A and NGC\,3801 will be undetectable in the radio beyond $z
\sim 0.04$ with the sensitivity of current instruments, these sources
provide one of the few means of obtaining more information about the
heating and compression of gas in more distant massive elliptical
galaxies, where such radio sources are likely to be present but
undetectable.

Throughout this paper, we adopt a cosmology with $H_{0} = 70$ km
s$^{-1}$ Mpc$^{-1}$, $\Omega_{M} = 0.3$ and $\Omega_{\Lambda} = 0.7$.
We adopt a luminosity distance fot NGC\,3801 of 52.6 Mpc, obtained by
correcting the heliocentric velocity of 3317 ks s$^{-1}$ \citep{lu93}
to the CMB frame of reference, which gives an angular scale of 1
arcsec = 0.25 kpc at the distance of NGC\,3801.

\section{Data analysis}

\subsection{Chandra}
We observed NGC\,3801 with {\it Chandra} for 60 ks on 2006 April 6.
The observation was taken in VFAINT mode to minimize the background
level. The data were reprocessed from the level 1 events file with
{\sc ciao} 3.3 and CALDB 3.2, including VFAINT cleaning. The latest
gain files were applied and the 0.5-pixel randomization removed using
standard techniques detailed in the {\sc ciao} on-line
documentation\footnote{http://asc.harvard.edu/ciao/}. In addition to
applying the standard good time intervals we carried out GTI filtering
using the {\it analyze\_ltcrv} script, which reduced the total exposure
time to 58,886 s.

\begin{figure}
\begin{center}
\plotone{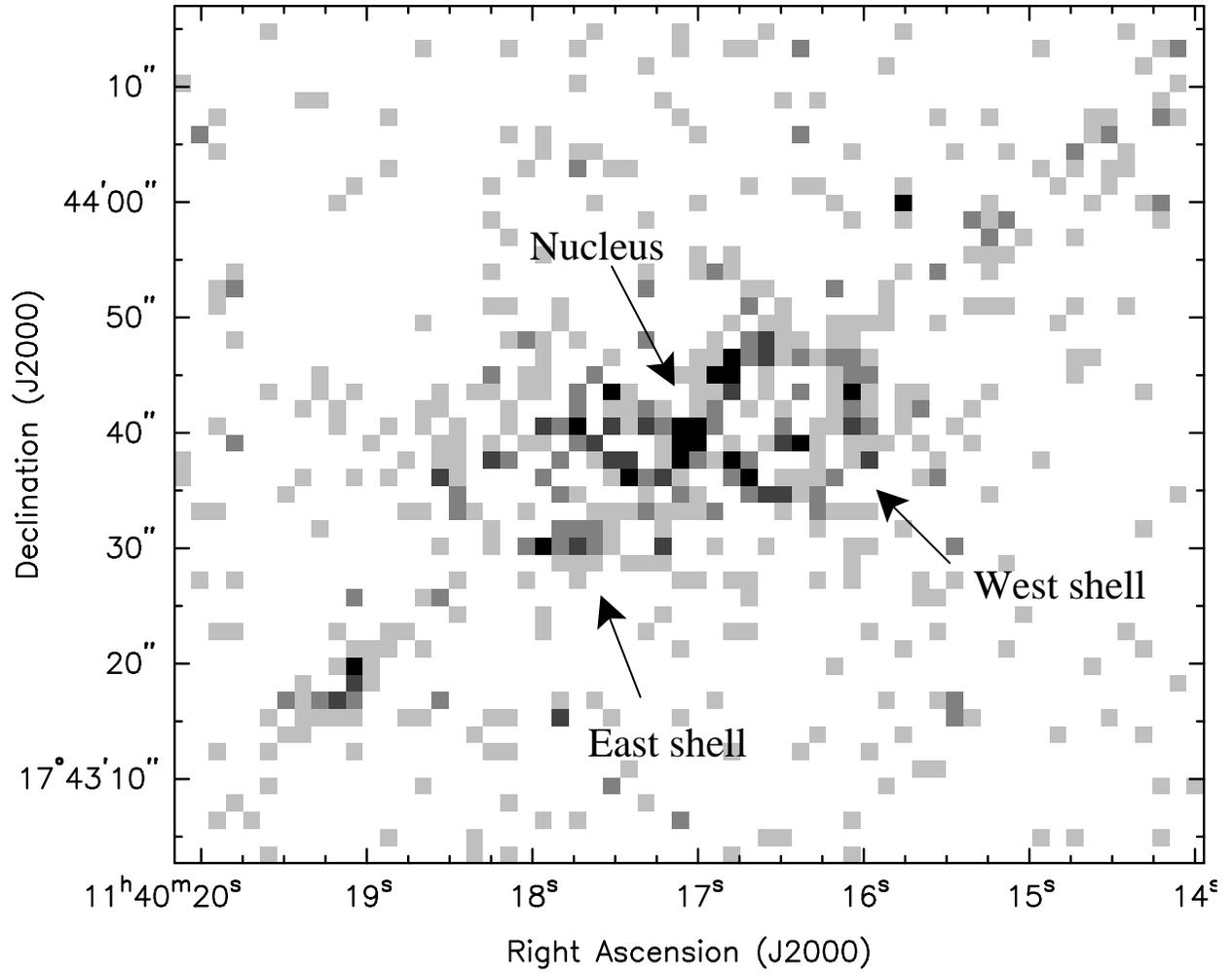}
\caption{A 0.5 -- 2 keV image of the {\it Chandra} data, binned by a
  factor of 3, revealing the symmetrical shell features associated
  with NGC\,3801}
\label{data}
\end{center}
\end{figure}

We produced a 0.5 -- 2 keV filtered image to examine the X-ray
emission associated with the radio galaxy, presented in
Fig.~\ref{data}. X-ray emission is detected from the core of
NGC\,3801, from regions surrounding the two radio lobes, indicated in
Fig.~\ref{data}, and extended emission from the galaxy atmosphere is
also detected. Fig.~\ref{overlay} shows a Gaussian smoothed image with
radio contours overlaid (see below) emphasizing the lobe-related X-ray
structure.

\begin{figure}
\begin{center}
\plottwo{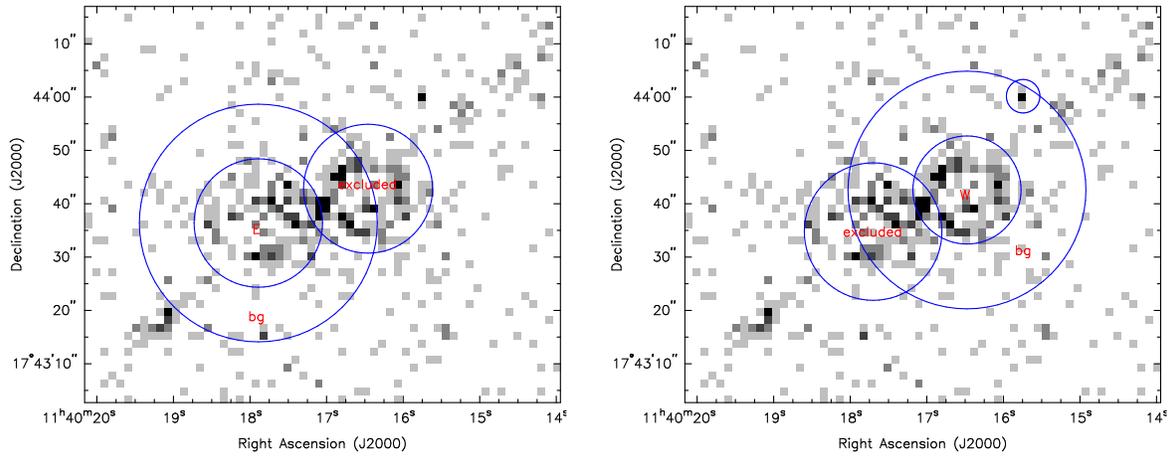}{f3b.eps}
\caption{The source and background regions used for spectral analysis of the X-ray shell
  emission.}
\label{regions}
\end{center}
\end{figure}

We extracted spectra from the core, the two lobes and the galaxy halo
using the {\it specextract} script, which also builds the appropriate
response files. Extraction regions are shown in Fig~\ref{regions}.
Local background regions adjacent to or surrounding the source
extraction regions were used. Spectra were grouped to 20 counts per
bin after background subtraction prior to spectral fitting, which was
carried out using {\sc xspec}. We assumed a fixed Galactic absorption
of $N_H = 2.3 \times 10^{20}$ cm$^{-2}$ \citep{dl90} in all of our
spectral fitting. 

\begin{figure}
\begin{center}
\plotone{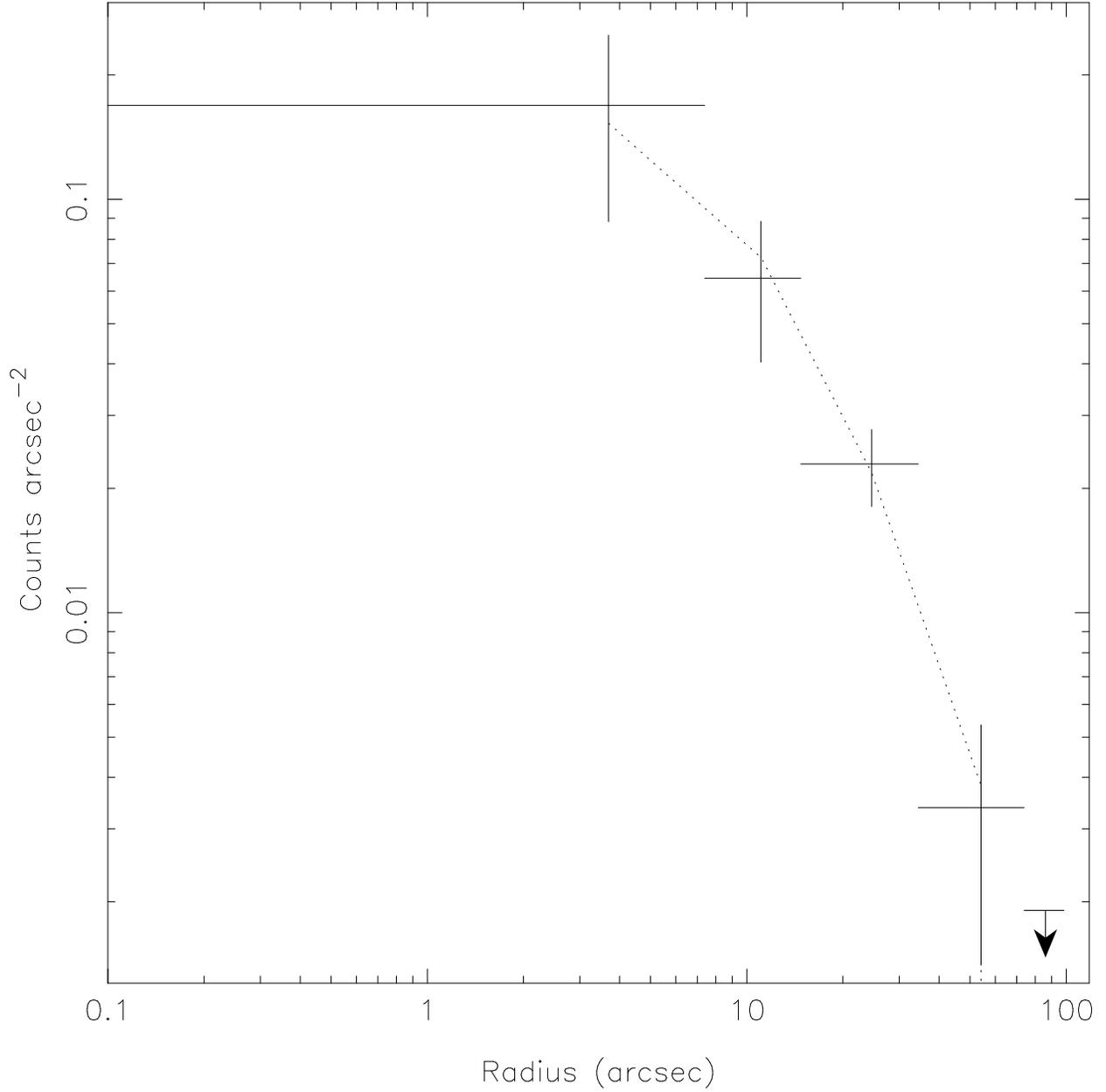}
\caption{Background-subtracted surface brightness profile extracted
  from the 0.5 -- 2 keV {\it Chandra} events list with best-fitting
  $\beta$ model having $\beta=0.45$ and $r_{c} = 7$ arcsec.}
\label{profile}
\end{center}
\end{figure}

We also extracted a radial surface brightness profile for NGC\,3801 to
constrain the X-ray emission from the ISM, masking out the radio lobe
regions and point sources, and using an outer annulus (between 74 and
123 arcsec) as a local background. The profile was extracted in the
0.5 -- 2 keV energy range, where the contribution from the nuclear
point source is negligible, so that the profile could be fitted by a
$\beta$ model convolved with the {\it Chandra} PSF, as parametrized by
Worrall et al. (2001). Further analysis of the radial profile is
described in Section~\ref{ism}. The background-subtracted radial
profile is shown in Fig.~\ref{profile}.

\subsection{Radio data}
We observed NGC\,3801 at 1.4 GHz and 4.9 GHz with the VLA on 2006
February 17 in A configuration, on 2006 June 27 in B configuration,
and on 2006 November 26 to supplement existing archival data. The data
were calibrated in the standard manner using {\sc aips}. The {\it
clean} algorithm was used to map the data for each frequency at each
configuration, and in each case self-calibration was carried out to
convergence using the clean model components. The final 1.4-GHz map
was obtained by first combining our A-configuration data with an an
earlier snapshot observation (AB920) in the same configuration to
improve our {\it uv} coverage. The archive data were independently
calibrated, and then the final {\it clean} model obtained from the new
data was used to calibrate the archive dataset, prior to combining the
{\it uv} data of the two observations using {\it dbcon}. The 1.4-GHz
B-configuration data were then calibrated using the best
A-configuration model, before being combined with the A-configuration
data with appropriate weighting, using {\it dbcon}. A final map, shown
in Fig.~\ref{overlay}, was then produced from the combined data. The
4.9-GHz data were reduced similarly, with the C-configuration data
first calibrated using the best B-configuration model before they were
combined. The final combined B and C configuration map at 4.9-GHz has
resolution matched to the combined A and B configuration map at
1.4-GHz, enabling spectral analysis to be carried out as discussed in
Section~\ref{age}.

\section{Results}

\subsection{Radio lobe related emission}
\label{lobes}
Although, the edge-brightned shell morphology of the lobe-related
emission immediately suggests a hot gas origin, we initially fitted
the spectra for the east and west lobe-related X-ray emission with
single power-law and {\it mekal} models with the aim of distinguishing
between a thermal and non-thermal origin. In neither case could a
single power-law model provide an acceptable fit to the data. We
calculate that the predicted X-ray flux from lobe inverse-Compton
scattering of CMB and synchrotron photons \citep[e.g.][]{c05b}, based
on the 1.4-GHz radio flux and assuming equipartition of energy density
in magnetic fields and particles, is at least two orders of magnitude
below that observed, so that we would not expect to see significant
emission from these processes. 

A good fit was obtained for the west lobe region with a {\it mekal}
model having $kT = 1.0^{+0.4}_{-0.3}$ keV, assuming $Z = 0.3$ solar,
giving $\chi^{2} = 2.5$ for 2 d.o.f. The unabsorbed 0.4 - 2 keV flux
from this component is $6.0^{+2.1}_{-1.5} \times 10^{-15}$ ergs
cm$^{-2}$ s$^{-1}$. For the eastern lobe we obtained a good fit
($\chi^{2} = 5.6$ for 3 d.o.f.) with $kT = 0.71^{+0.09}_{-0.04}$ keV,
giving an unabsorbed 0.4 - 2.0 keV flux of $(6.9\pm0.8) \times
10^{-15}$ ergs cm$^{-2}$. We carried out a free abundance fit for the
west lobe, which did not significantly alter the measured temperature,
and for the east lobe (which had too few bins for a similar test) we
verified that varying the abundance between 0.28 (the lowest
acceptable value) and 0.5 solar did not significantly affect the
temperature, though in both cases the {\it mekal} normalization is
affected (we discuss the effect of this uncertainty on the inferred
gas density in Section~\ref{physical}). We also compared a range of
background regions, including a large off source region, and regions
on either side of the radio lobes, none of which affected the
temperature determination for each lobe. Finally, we tested a combined
power-law + mekal model for the west lobe, but obtained a very low
power-law normalization and no significant change in temperature or
temperature uncertainty. As stated above, there is no plausible
physical origin for non-thermal X-ray emission of this strength
associated with the radio lobes, and so we do not consider non-thermal
models further.

\begin{figure}
\begin{center}
\plottwo{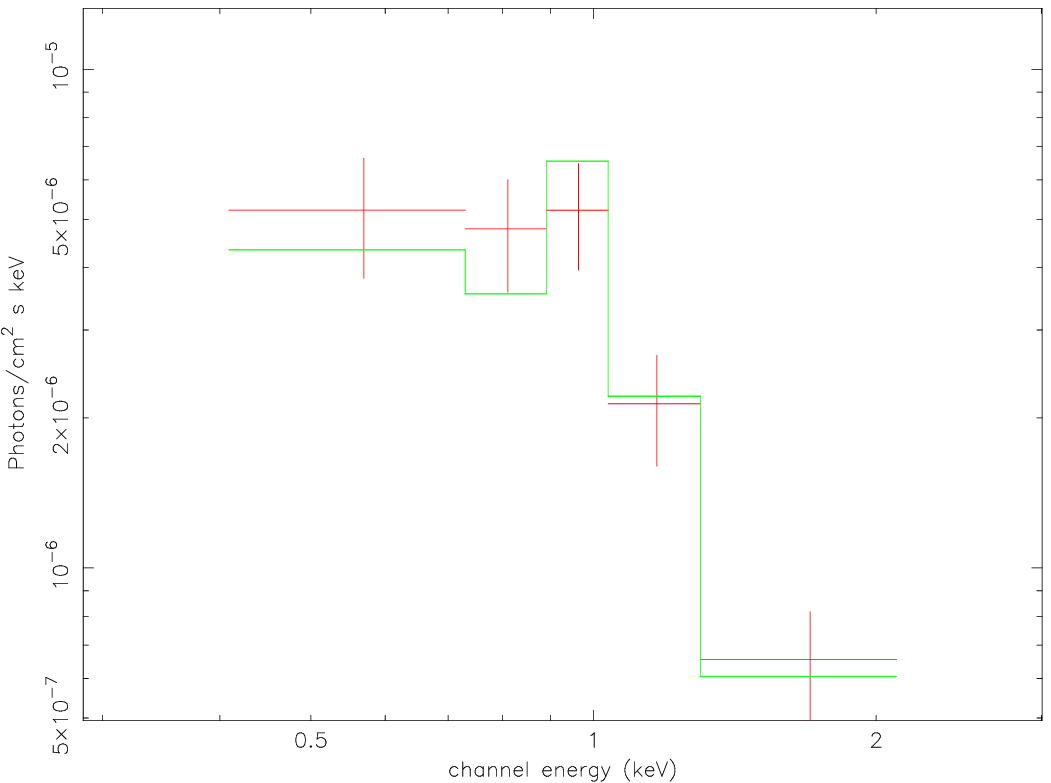}{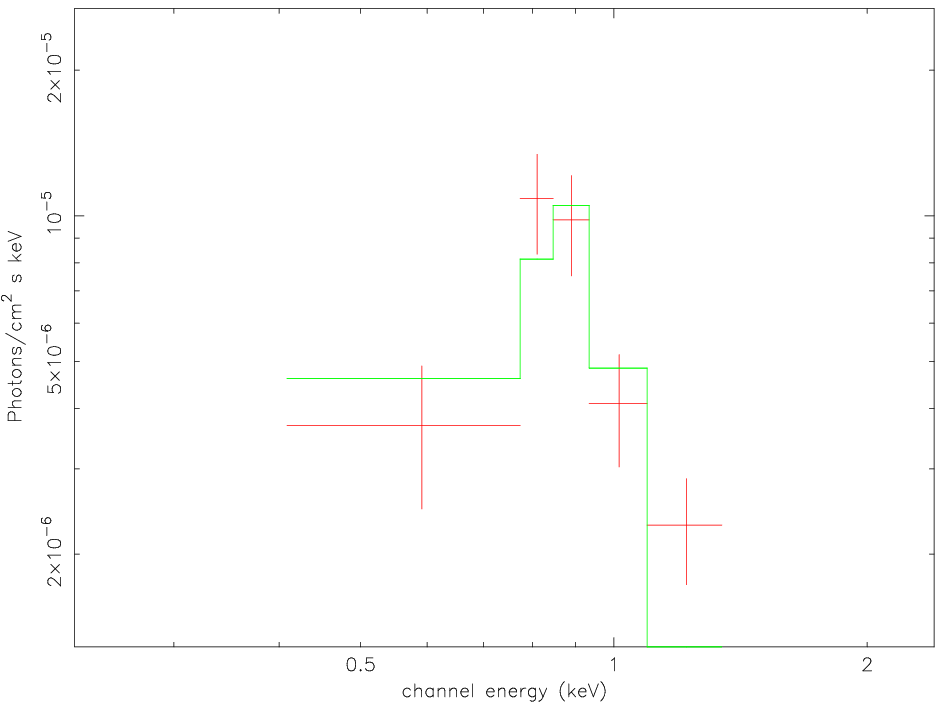}
\caption{The spectra for the west and east lobe-related regions of
  X-ray emission with best-fitting {\it mekal} models as described in
  the text.}
\label{lobespec}
\end{center}
\end{figure}

If the hot gas is due to shocks, it is likely that the shells will
have temperature structure, so that a single temperature {\it mekal}
model is not ideal. In addition, the dust lane structure of the host
galaxy could lead to variable absorption across the eastern shell.
Unfortunately the photon statistics do not allow us to test for either
of these scenarios and so in the discussion that follows we adopt the
physical parameters obtained from the single temperature {\it mekal}
fits. Our adopted spectral models for the lobe-related X-ray emission
are shown in Fig.~\ref{lobespec}.

\subsection{Interstellar medium and X-ray binary population}
\label{ism}
We used surface brightness profile analysis to investigate the
extended emission from the X-ray halo of NGC\,3801, using the 0.4 -- 1
keV energy range and masking point sources to minimize any
contamination from the resolved X-ray binary population in this
galaxy. Extended emission is detected to a radius of $\sim 75$ arcsec,
which corresponds to $\sim 17$ kpc. As the contribution from the
central AGN in this energy range is negligible due to its heavy
absorption, we simply fitted a $\beta$ model convolved with the {\it
Chandra} PSF using the model of \citet{wor01}. We obtained
best-fitting values of $\beta = 0.45$ and $r_{c} = 7$ arcsec ;
however, both the slope and core radius are poorly constrained due to
the small numbers of counts. The profile and best-fitting model are
shown in Fig.~\ref{profile}. The extent of the detected gas halo is
comparable to the extent of stellar light detected in the ground-based
observations of \citet{pel90}, who calculated an effective radius of
11.1 kpc (transformed to our luminosity distance) based on the
measurements of \citet{bur87}; however, the $R$-band surface
brightness profile of \citet{pel90} shows that the stellar
distribution is considerably steeper than that of the hot gas.

\begin{figure}
\begin{center}
\plotone{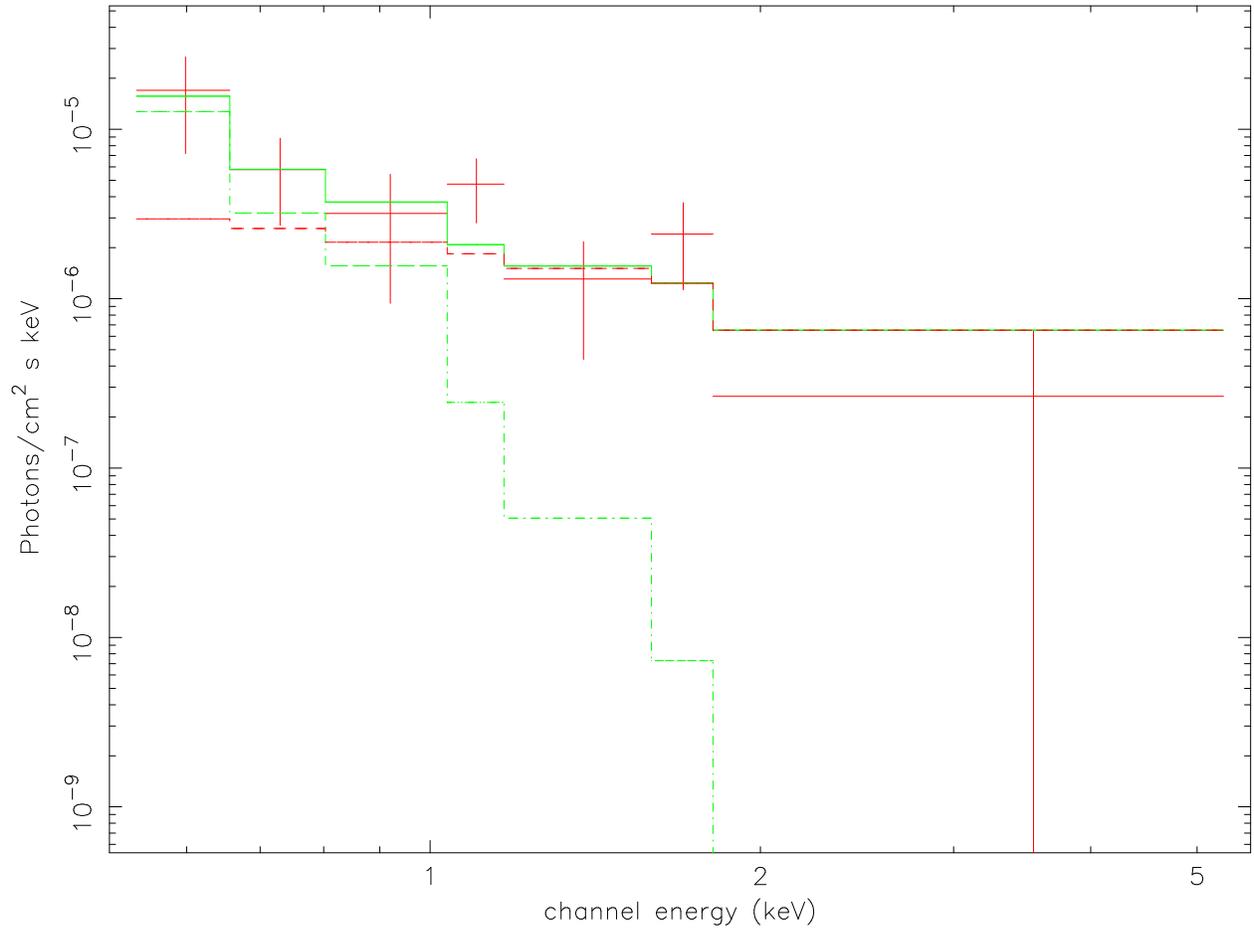}
\caption{The spectrum of the ISM with best-fitting {\it mekal} plus
  power-law model, as discussed in detail in the text.}
\label{ismspec}
\end{center}
\end{figure}

We then extracted a spectrum for the region sampled by the radial
profile (excluding the nucleus and lobe regions). A single {\it mekal}
model resulted in an implausibly high temperature; however, there is
likely to be a significant contribution from the unresolved X-ray
binary population and so we fitted a model consisting of a {\it mekal}
plus flat ($\Gamma = 1.0$) power-law to account for this component. A
best-fitting temperature of $0.23^{+0.21}_{-0.09}$ keV was obtained
(for fixed abundance of 0.4 solar, the measured value for Centaurus A,
Kraft et al. 2003) resulting in a value of $\chi^{2} = 3.8$ for 4
d.o.f. Fig.~\ref{ismspec} shows the ISM spectrum with this
best-fitting model. The temperature is not sensitive to abundance;
however, the mekal normalisation varies by a factor of $\sim 8$ if the
abundance is varied between 0.l and 1.0 times solar. We carried out
simulations using the {\it fakeit} command in {\sc xspec}, which
demonstrated that we can obtain an unbiased estimate of the
temperature for a value of 0.23 keV in the abundance range 0.1 to 1.0
times solar, with a 90 percent confidence range similar to the quoted
uncertainties for an abundance of 0.4 times solar. Our fitted
temperature is also in reasonable agreement with that inferred from
the galaxy velocity dispersion (see Section~\ref{energy}). We measured
an unabsorbed 0.4 - 7.0 keV flux of $4^{+5}_{-2} \times 10^{-15}$ ergs
cm$^{-2}$ s$^{-1}$ from the thermal component, and a 1-keV flux
density of $1.3\pm0.5$ nJy from the PL component. Modelling the hard
component with a thermal bremsstrahlung component of $kT = 5 - 10$
keV, as might be expected for a population of low-mass X-ray binaries,
does not significantly affect the best-fitting temperature or the
luminosity of the {\it mekal} component. We determined the bolometric
X-ray luminosity of the ISM, by integrating the best-fitting $\beta$
model and using the best-fitting thermal model to convert from the
total number of counts, to be $\sim 2 \times 10^{40}$ ergs s$^{-1}$.
Since NGC\,3801 has $M_{B} = -21.33$ \citep{pel90}, this puts it in
the middle of the $L_{X}/L_{bol}$ distribution for elliptical
galaxies, as studied by \citet{osul03}. For an abundance of 0.4 times
solar, the central density of the gas is $3 \times 10^{-2}$ cm$^{-3}$
(ranging from $2.5 \times 10^{-2} - 6 \times 10^{-2}$ cm$^{-3}$ for
abundances ranging from 0.1 to 1.0 times solar). We note that the
radial distribution of the gas may differ somewhat from the beta model
fitted above, since emission from X-ray binaries contributes $\sim 40$
percent of the flux in the energy range used; however, our X-ray
surface brightness profile is much flatter than the stellar
distribution \citep{pel90}, so that it is likely to be dominated by
the gas emission. Any systematic effect of the X-ray binary
contribution must be to reduce the gas density at the radii of
interest discussed further in Section~\ref{physical}. The unabsorbed
0.4 - 7 keV luminosity of the unresolved X-ray binary component is
$\sim 7 \times 10^{39}$ erg s$^{-1}$, which is comparable to the total
X-ray binary luminosity measured by Kraft et al. (2001) for Centaurus
A.

There are several bright point sources within the $D_{25}$ radius of
the galaxy. To determine if any of these are likely to be
ultra-luminous X-ray binaries (ULXs) related to NGC 3801, we
determined the positions and fluxes of the sources using the CIAO tool
{\it wavdetect}. The brightest source contains 28 counts in the
0.5-2.0 keV band. To convert count rate to luminosity, we assume a 5
keV bremsstrahlung spectrum with Galactic absorption. A count rate of
10$^{-3}$ cts s$^{-1}$ in the 0.5-2.0 keV band corresponds to a flux
of $6.3 \times 10^{-15}$ ergs cm$^{-2}$ s$^{-1}$ in the 0.1-10.0 keV
band (unabsorbed). The observed rate from the brightest source
corresponds to a luminosity of 1.4$\times$10$^{39}$ ergs s$^{-1}$.
Based on the log(N)-log(S) relation of Tozzi et al. (2001), we expect
$\sim$0.5 background AGN of this flux or higher within the $D_{25}$
radius. There are eight other sources detected in the $D_{25}$ radius
with fluxes a factor of 3 to 7 lower than the brightest source. The
X-ray luminosity of the brightest source does not qualify it as a ULX
if it is related to NGC 3801, and there is a significant probability
that it is an unrelated background AGN. Some of the less luminous
sources are likely to be LMXBs related to the host galaxy at the high
end of the luminosity function, none of which are ULXs. We conclude
that none of the point sources are ULXs.

\subsection{Core}
\label{cores}
The nuclear spectrum of NGC\,3801 was poorly fitted with a single
power-law model with Galactic absorption ($\chi^{2}/n > 2$): extremely
prominent residuals are present at hard energies. We therefore added a
second model component consisting of a heavily absorbed power law,
which resulted in a good fit. A similar two power-law model was found
to provide the best description of the nuclear spectrum of Centaurus A
\citep[e.g.][]{ev04}, and is commonly used to fit spectra of
narrow-line radio galaxies.

\begin{figure}
\begin{center}
\plotone{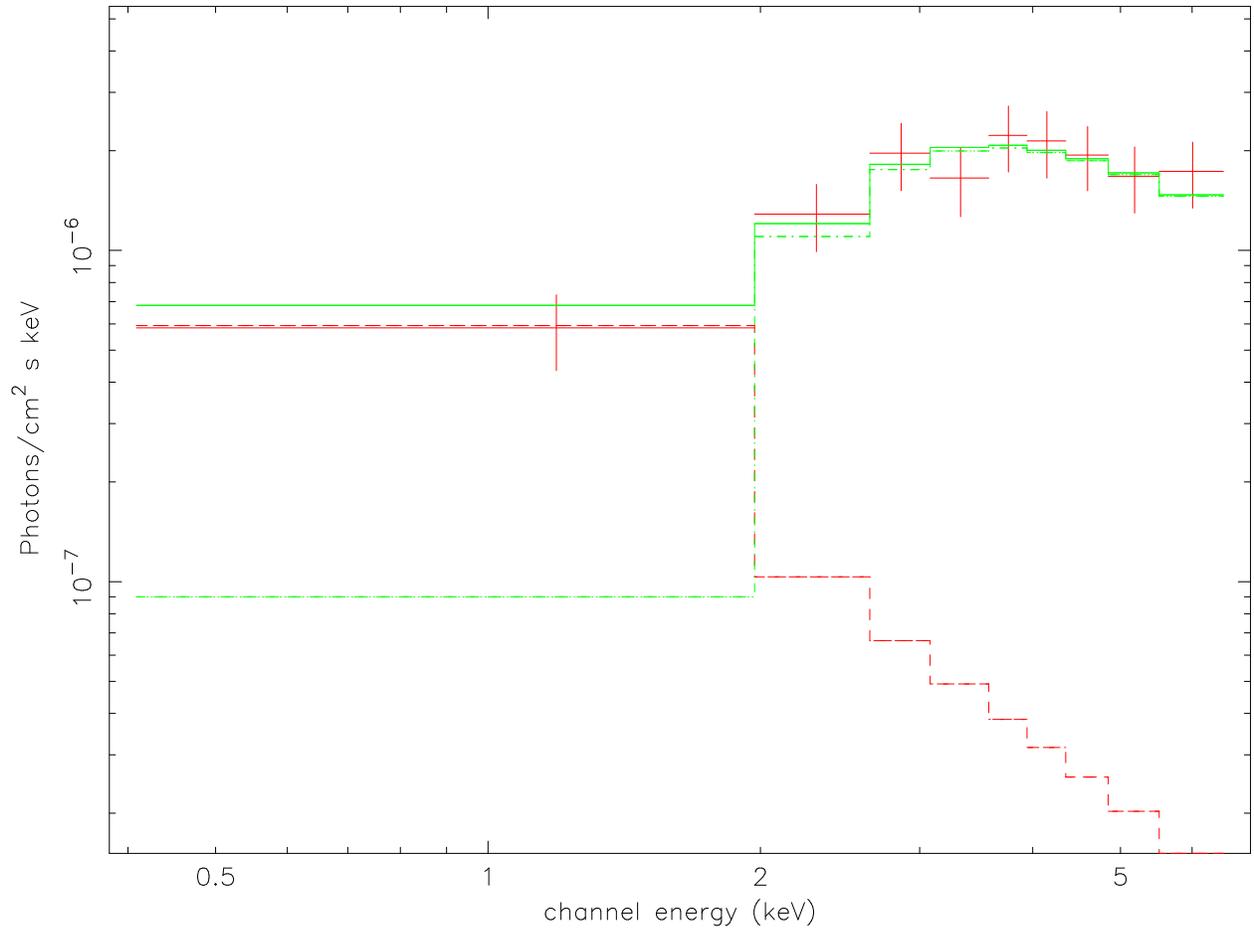}
\caption{Core spectrum (in the energy range 0.4 - 7 keV) with
best-fitting model as described in the text.}
\label{core}
\end{center}
\end{figure}

The power-law index of the unabsorbed component is poorly constrained,
and so we fixed its value at $\Gamma=2$ (a typical value for
unabsorbed power law components in FRI radio galaxies, e.g.
\citealt{ev06}). The best fitting values of $\Gamma$ and $N_H$ for the
second component were $0.6^{+0.4}_{-0.5}$ and $3.1^{+2.6}_{-1.6}
\times 10^{22}$ cm$^{-2}$, respectively, giving $\chi^{2} = 1.4$ for 5
d.o.f. We then fixed the second power-law index at the more physically
plausible value of 1.5 \citep[e.g.][]{ev06,hec06}, which gave a
best-fitting value of $N_{H} = 6.4^{+1.3}_{-1.1} \times 10^{22}$
cm$^{-2}$ and $\chi^{2} = 2.2$ for 6 d.o.f. This best-fitting model is
shown in Fig.~\ref{core}. The unabsorbed 1-keV flux densities in this
model were $0.3\pm0.1$ nJy and $16\pm2$ nJy for the first and second
power-law components, respectively, corresponding to an unabsorbed
bolometric luminosity from the obscured component of $3 \times
10^{41}$ ergs s$^{-1}$. The 5-GHz flux density of the core is 3.5 mJy,
which means that the primary (unabsorbed) power law component lies
close to the radio-X-ray core correlation for components with low
intrinsic absorption reported by \citet{ev06}, whereas the second
(absorbed) power law lies significantly above the correlation, as is
generally the case for heavily absorbed components. We therefore
interpret the primary power law as jet-related in origin, and the
second, heavily absorbed power law as having an accretion-related
origin.

We used archive HST data (taken using the F555W and F814W filters on
WFPC2) to measure the extinction from the central dust lane, which was
converted to $E(B-V)$ and subsequently column density of neutral
hydrogen using the relations of \citet{zom}. We obtained a value of
$N_{H} \sim 2 \times 10^{21}$ cm$^{-2}$, which is significantly lower
than the column density obtained from our spectral fit, so that we
conclude that, in addition to the absorption from the dust lane, the
NGC\,3801 must contain a substantial quantity of absorbing material
instrinsic to the nucleus. Heavily absorbed nuclear spectra are not
usually found in FRI radio galaxies (e.g. Evans et al. 2005); but
Centaurus A is an exception, and it would appear that NGC\,3801 is a
second example of an FRI radio galaxy with a nucleus having high intrinsic
absorption. We discuss the implications of this result further in
Section~\ref{hostgal}.

\section{Radio source age}
\label{age}
We can use our two-frequency radio data to give us some constraints on
the age of the radio source. The two-point spectral indices between
1.4 and 4.9 GHz in the outer regions of the source are only marginally
steeper than those in the inner regions, as expected if the radio
source is young. We fitted a Jaffe \& Perola (1973) aged electron
spectrum to the radio data in the outer regions of the source,
assuming a constant ageing magnetic field of 2 nT (the mean
equipartition field in the lobes). The main uncertainty in this model
is the value of the low-frequency electron energy index (the
`injection index'). Young \etal\ (2005) have recently argued that in
normal FRI sources the low-energy electron energy index tends to be
close to 2.1 ($\alpha = 0.55$, where we define $\alpha$ in the sense
that $S \propto \nu^{-\alpha}$). If we adopt this injection index then
the spectral index in the steepest part of the lobe ($\alpha \approx
0.69$) corresponds to a source age of $2.4 \times 10^6$ years. If the
injection index were steeper, the age would be reduced: for example,
for an injection index corresponding to $\alpha = 0.65$, which is the
flattest spectrum we observe in the jets, we fit an age of $8 \times
10^5$ years. It is probably reasonable to treat the age of $2.4 \times
10^6$ years as an upper limit. This value is in good agreement with
the age we obtain from source dynamical arguments in
Section~\ref{dynamics}.

\section{Discussion}

\subsection{Physical properties of the gas shells, ISM and radio lobes}
\label{physical}
The spectral analysis presented in Section~\ref{lobes} revealed that
the lobe-related thermal emission is significantly hotter than the
surrounding ISM gas for both lobes. In this section we therefore
investigate further the plausibility of a model in which we are
observing shock-heated shells of gas around the radio lobes of
NGC\,3801.

We calculated the density and pressure of the gas in each shell region
by assuming a spherical shell of gas with an outer radius defined by
the outer edge of the detected X-ray emission ($R_{out} = 8$ arcsec
for both lobes). We assume that the radio source is close to the plane
of the sky, as suggested by Das et al. (2005) based on the
two-sidedness of the jet and the orientation of the nuclear disk, and
so we used projected distances to calculate the shell volume (if the
source were at 45 degrees to the angle of sight, we would be
underestimating the shell volumes by roughly a factor of 2). It is
difficult to estimate the thickness of the shells accurately, because
the photon statistics are poor, and so we chose conservative upper and
lower limits based on the apparent thickness of the emission in the
{\it Chandra} data (corresponding to a projected distance of $\sim
500$ pc), and the modelled thickness obtained for the shell in
Centaurus A where the data quality is sufficient to allow a more
accurate determination (corresponding to $\sim 200$ pc). The gas
densities and pressures in the shells were then determined from the
best-fitting {\it mekal} models for each lobe, and are given in
Table~\ref{properties} as a range based on the two volume estimates.
The uncertainty due to the volume completely dominates over the
statistical uncertainties on gas temperature and abundance. The
structure of the shells appears clumpy in the X-ray images; however,
there are too few counts to be able to determine adequately
constrained physical properties for sub-regions of the shells. It is
necessary to bear in mind in the discussion that follows that the
density and pressure values we use for the shells are average values:
the distribution of X-ray counts suggests that the surface brightness
and hence the gas density in the shells is higher towards the centre
of the radio source. We note that the contribution of X-ray binary
emission to the surface brightness profile (Fig.~\ref{profile}) is
unlikely to affect significantly our inferred densities, and any small
effect will be in the direction of reducing the ISM density. The
conclusions of Section~\ref{dynamics} are therefore unaffected.

Since the external gas density and pressure in the environment of the
radio source varies around the edge of the shells, we calculated the
external density and pressure at several locations around the edge of
the shells. Table~\ref{properties} lists these properties at a
distance of 10 arcsec from the nucleus of NGC\,3801, which roughly
corresponds to the midpoint of each shell. The uncertainty on the ISM
density is dominated by including a systematic uncertainty due to the
unconstrained metal abundance. In Section~\ref{dynamics}, we consider
how the variation of shell and ISM properties may affect our results.

We also calculated the equipartition minimum internal pressure of the
radio lobes, using measurements of the 1.4-GHz flux density for each
lobe to normalize the synchrotron spectrum. We assumed a broken
power-law electron distribution with initial electron energy index,
$\delta$, of 2.1, $\gamma_{{\rm min}} = 10$ and $\gamma_{{\rm max}} =
10^{5}$, and a break at $\gamma_{{\rm break}} = 10^{3}$, and modelled
the lobes as spheres having volumes of $2.1 \times 10^{10}$ pc$^{3}$
and $2.5 \times 10^{10}$ pc$^{3}$ for the west and east lobes,
respectively. The resulting pressures are given in
Table~\ref{properties}.

The mass in each shell was also calculated in order to compare with
the mass expected to have been swept-up by the expanding radio lobes.
We obtained a mass of $(5.3 - 7.7) \times 10^{6}$ M$_{\sun}$ for the
West shell and $(2.0 - 6.1) \times 10^{6}$ M$_{\sun}$ for the East
shell. The ISM mass swept up was calculated by modelling the radio
lobes as expanding cones of 45$^{\circ}$ half-opening angle and
integrating the best-fitting gas density distribution out to a
distance of 2.1 kpc, which gives a total mass of $\sim 1.7 \times
10^{7}$ M$_{\sun}$, in fairly good agreement with the total shell
mass. These calculations are therefore consistent with a scenario in
which the shells consist of compressed material from the regions of
the ISM through which the radio source has passed.


\subsection{Dynamics of the lobe/ISM interaction}
\label{dynamics}
As shown in Table~\ref{properties}, the pressures in the hot shells of
gas are higher than the minimum internal pressure of the radio lobes;
since the shells and radio lobes must be in pressure balance, this
suggests that the true radio lobe pressure is a factor of $\sim 4-6$
times the minimum value. Such a departure from minimum energy is
generally observed in FRI radio galaxies for which external pressures
can be measured \citep[e.g.][]{mor88,c03} and is also seen in the
inner lobes of Centaurus A \citep[e.g.][]{kra03}. NGC\,3801 is one of
the few known FRI radio galaxies for which the {\it minimum} internal
pressure is higher than the external pressure in the ISM (see
Table~\ref{properties}); this is also the case for Cen A, and was one
of the reasons we selected NGC\,3801 as a {\it Chandra} target. The
pressure in the X-ray shells of NGC\,3801 (and by inference the true
pressure in the radio lobes) is more than an order of magnitude higher
than the local pressure in the interstellar medium.

The density contrast between the X-ray shells and the surrounding,
undisturbed gas is slightly higher, but consistent with a value of
$\sim 4$ for both the east and west shells, using the average shell
density determined from the shell spectra, and the external gas
density roughly halfway along the shells. This is the value expected
if we are observing a strong shock for which the Rankine-Hugoniot jump
conditions hold. 

If we compare our value for the mean shell density with the external
density at a distance of 5 arcsec, corresponding to the inner parts of
the shells, the density contrast is $\sim 3$; however, as shown in
Figs.~\ref{data} and~\ref{overlay}, the brightest parts of the shell
are in this region, so that the shell densities will be significantly
higher than the mean value. The true density constrast in these inner
regions is likely to be consistent with the expected value. The outer
regions of the shell are fainter, so that the shell density is likely
to be lower here. We conclude that it is plausible that the density
contrast is in agreement with the value of 4 expected for a strong
shock around the entire perimeter of the shells. Since the shell
pressure should correspond to the internal radio lobe pressure, it
should be uniform. This implies that the shell temperature at its
outer edge should be considerably higher than the mean temperature in
order to compensate for the lower gas density. Assuming a density
contrast of 4, we calculate a temperature of $\sim 2.5$ keV for the
outer edge of the shell. The contribution of X-ray emission from this
hotter region is not expected to be significant due to the lower
density.

Assuming a density contrast of 4, we can use the Rankine-Hugoniot
conditions to estimate the Mach number of the shock, as follows
(Landau \& Lifshitz 1990):
\begin{equation}
{\cal M} = \sqrt{\frac{4(\gamma + 1)\frac{T_{shell}}{T_{ism}} + \gamma
    - 1}{2\gamma}}
\end{equation}
which gives Mach numbers of $3.8\pm1.3$ and $3.2\pm1.0$ for the
west and east shells respectively. The measured temperature increases
are therefore consistent with the interpretation of both shells as
originating in the boundary layer of a strong shock.

We also used a simple ram pressure balance calculation as a second
means of estimating the shock advance speed as follows (using a
density constrast of 4, consistent with the observed gas properties of
the shell and ISM):
\begin{equation}
P_{shell} + \frac{1}{4}\rho_{ism} v_{ism}^{2} = P_{ism} + \rho_{ism} v^{2}
\end{equation} 
This leads to the following expression for ${\cal M}_{shell}$, the
Mach number of the expanding shocked shell:
\begin{equation}
{\cal M}_{shell} = \frac{2\sqrt{5}}{3}\sqrt{\frac{P_{shell} - P_{ism}}{P_{ism}}}
\end{equation}
which corresponds to ${\cal M}$ in the range $3.5 - 8.1$ for the West
shell and in the range $2.9 - 6.8$ for the East shell, taking into
account the uncertainty in the shell volume and the uncertainty in the
ISM density due to the unconstrained abundance. These estimates are
consistent with the values obtained from the measured temperature
jump.

We conclude that we have indeed detected shells of strongly shocked
gas around both radio lobes of NGC\,3801, with Mach numbers of $3 -
8$, and shock conditions roughly consistent with the Rankine-Hugoniot
jump conditions). If we assume that the radio source has expanded at
${\cal M} \sim 4$ throughout its lifetime, this gives a source age of
$\sim 2 \times 10^{6}$ years, which is in good agreement with the
upper limit estimated from spectral ageing in Section~\ref{age}. The
sound speed in the ISM is $\sim 211$ m s$^{-1}$, so that a Mach number
of 4 corresponds to a lobe expansion speed of $\sim 850$ km s$^{-1}$.

There are a number of differences between the results we obtain for
NGC\,3801 and the shell in Cen A (Kraft et al. 2003). Firstly, the
shell morphologies are different: in Cen A, only one shell is detected
(although there may be some much fainter shell-like emission
associated with the north-eastern lobe). In addition, the shell in Cen
A is brightest at the outer edge of the radio lobe, whereas the
NGC\,3801 shells are brightest in the regions towards the nucleus. If
lobe expansion is supersonic everywhere, the expansion of a uniform
pressure radio lobe in an isothermal $\beta$-model atmosphere should
mean that the shock is weakest close to the nucleus; however, the
higher density of the ISM in this region means that after compression
its X-ray emission would nevertheless be expected to dominate. It is
not clear why this is not observed in Cen A.

\subsection{Implications for ISM energetics}
\label{energy}
The total thermal energy stored in the detected hot gas shells is
calculated to be $(4.6\pm2.3) \times 10^{55}$ ergs and $(3.3\pm0.6)
\times 10^{55}$ ergs for the West and East shells, respectively, or
$\sim 8 \times 10^{55}$ ergs in total. Assuming a Mach number of 4,
the kinetic energy of the shells is $(4.7\pm2.4) \times 10^{55}$ ergs
per shell. Therefore, the total energy stored in the shocked shells is
($1.7\pm0.4) \times 10^{56}$ ergs. This is good agreement with the
total work available from the radio lobes (assuming that the true
internal lobe pressure is that of the shocked shells), $P_{int}V \sim
1.3 \times 10^{56}$ ergs. We calculate that the energy required to
inflate both lobe cavities is $\sim 7 \times 10^{54}$ ergs, a factor
of $\sim 25$ lower than the energy stored in the shells. Shock heating
is therefore the dominant energy transfer mechanism during this stage
of the radio source's evolution. Using the best-fitting gas density
distribution, we find that this corresponds to the total thermal
energy in the ISM within $\sim 11$ kpc (and roughly 25 percent of the
total thermal energy within 30 kpc). The radio-source shock heating
detected in NGC\,3801 is therefore having a significant energetic
impact on the host galaxy's ISM, and may provide enough energy to heat
the gas sufficiently for the entire ISM to be expelled from the
galaxy. We note that in addition, the internal energy of the radio
source ($\sim 4 \times 10^{56}$ ergs) must eventually be transferred
to the environment, although over a much longer timescale, and
possibly at much larger distances.

Assuming a source age of $2 \times 10^{6}$ years (Section~\ref{age}),
this means that the radio source is putting work into the ISM at a
rate of $3 \times 10^{42}$ ergs s$^{-1}$, which should correspond to a
significant fraction of the jet power. This is a plausible value,
given that the most powerful FRI jets such as that in 3C\,31 have jet
powers of $\sim 10^{44}$ ergs s$^{-1}$ \citep[e.g.][]{lai02}. Our
measurements of jet impact have therefore provided an independent
means of estimating radio-jet power, which supports the more general
validity of using jet-power arguments to estimate the energy input
from radio galaxies into their environments. The rate of mechanical
work that we calculate is roughly an order of magnitude higher than
the X-ray luminosity of the obscured nuclear component
(Section~\ref{cores}), so that the AGN must be more efficiently
converting energy into jet production than into radiation.

To investigate the relationship between this energy output and
accretion on the central black hole, we used the stellar velocity
dispersion ($\sigma = 168$ km s$^{-1}$, \citealt{pel90}) to estimate a
black hole mass of $\log(M_{BH}/M_{\sun}) = 8.43$, using the relation
of \citet{tre02}. Assuming an efficiency of $\eta = 0.1$, the energy
output we have measured from the radio source and shocked gas shells
would require a total accreted mass of $\sim 800$ M$_{\sun}$, or an
accretion rate of $\sim 4 \times 10^{-4}$ M$_{\sun}$ yr$^{-1}$, given
the source age estimate above. For comparison, we calculate a Bondi
accretion rate of $2.5 \times 10^{-3}$ M$_{\sun}$ yr$^{-1}$ from the
central hot gas, which suggests that Bondi accretion could be
sufficient to provide the energy observed to be being transferred to
the ISM from the central AGN by the radio lobes if the efficiency is
relatively high \citep[cf.][]{allen}.

\section{An extended, linear feature aligned with the radio jet}

\begin{figure}
\begin{center}
\plotone{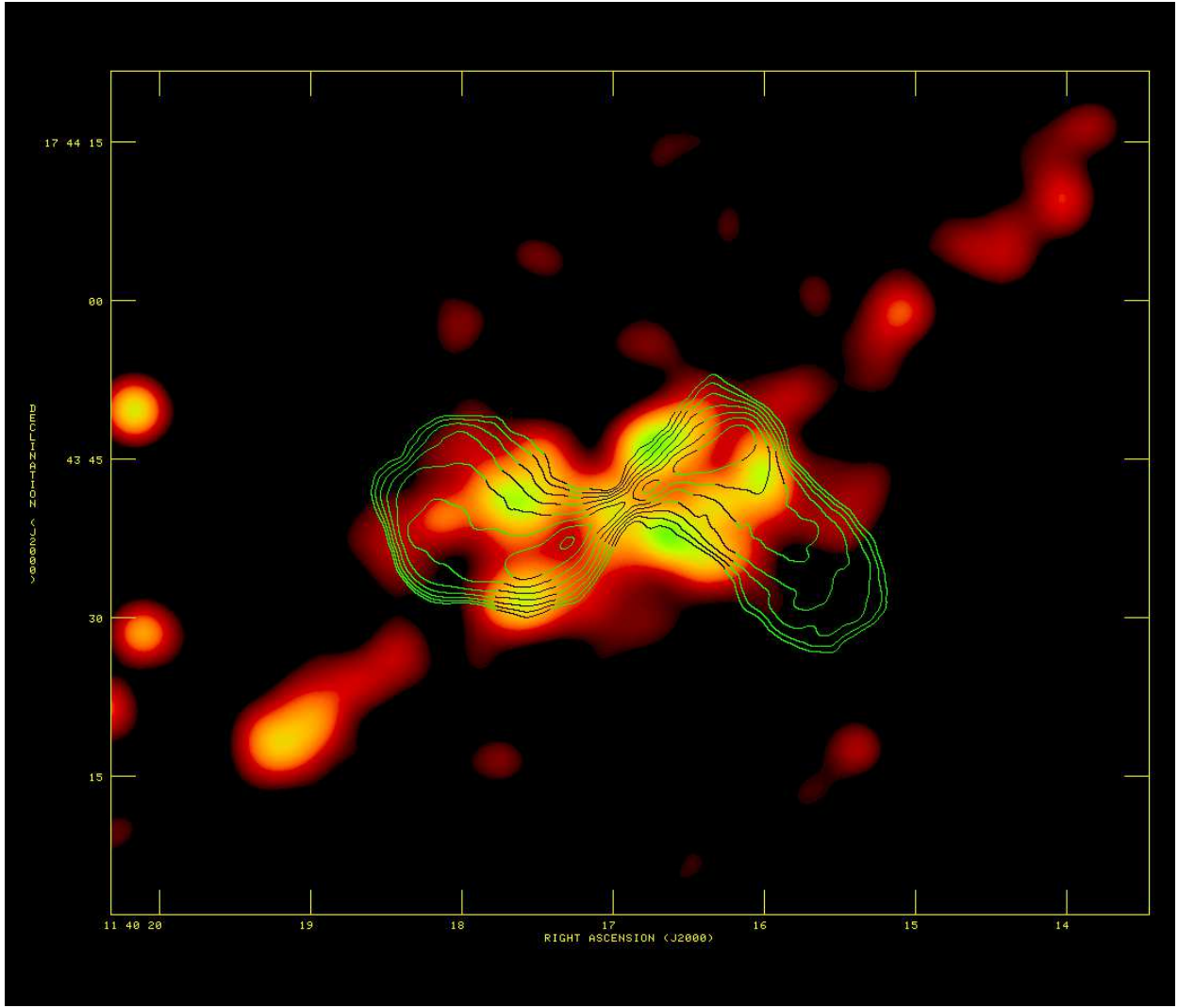}
\caption{Gaussian smoothed ($\sigma = 1.97$ arcsec) 0.4 -- 1 keV image
  of the {\it Chandra} data, with 1.4-GHz radio contours overlaid,
  indicating the soft, extended feature aligned with the jet axis.}
\label{streak}
\end{center}
\end{figure}

In addition to the bright shells discussed in the previous sections,
the {\it Chandra} data also reveals a linear feature on scales of 10
-- 20 kpc (shown in the 0.4 -- 1 keV image of Fig.~\ref{streak}),
which appears to be well-aligned with the inner radio jet axis. This
feature has an extremely soft spectrum, well-fitted by a power law of
photon index $\Gamma \sim 3.6$. The spectrum cannot be fitted by a
thermal model with $kT > 0.1$ keV. There is no radio emission
associated with this feature, nor does it appear to be associated with
any optical or emission-line features. Although it has a similar
orientation to one of the dust lanes, their position angles differ by
$\sim 15$ degrees. The X-ray feature is not aligned with the chip
axis, so that it is unlikely to be an instrumental artefact. We
believe that the most plausible origin for this soft X-ray emission is
a tidal or ram-pressure stripped tail, related to the
merging/interactions of NGC\,3801 with other nearby galaxies,
including NGC\,3802 to the north. The X-ray emission is therefore
likely to be thermal emission from cooler, dense material stripped
from the ISM of galaxies interacting with NGC\,3801. Such emission
could mimic the steep power-law spectrum we observe. 

It could be argued that the presence of X-ray features produced by
tidal debris might suggest a tidal origin for the features coincident
with the radio lobes. It is not possible to rule out such a model
definitively; however, the strong morphological agreement between
radio and X-ray structure, the significant differences in the spectral
properties of the shell features and tidal tails, and the fact that we
know the radio lobes to be overpressured and hence supersonically
expanding, all support our interpretation of the radio-related X-ray
emission as shocked gas.

\section{Relationship between the host galaxy and radio-source properties}
\label{hostgal}
Both Cen A and NGC\,3801 possess strong evidence for a recent merger
with a gas-rich galaxy, which may be more typical of FRII host
galaxies compared to the hosts of more powerful FRIs
\citep[e.g.][]{hec86,col95}. It is therefore interesting to consider
the relationship between their merger history, radio outbursts, and
their effects on the ISM as revealed by {\it Chandra}. NGC\,3801 and
Cen A have another feature in common, which may be related to their
merger history: they are the only two FRI radio galaxies that we know
of whose nuclear spectra require the presence of a high absorbing
column ($N_{H} \gtrsim 5 \times 10^{22}$), so that they are the only
known FRIs that possess any evidence for the torus often found in
powerful, high-excitation (FRII) radio galaxies and quasars
\citep{hec06}. Although we cannot draw strong conclusions from two
objects, it is interesting to speculate that these two sources may be
fuelled in the same way as FRII radio galaxies. It is possible that
these two radio galaxies represent a class of post-merger system in
which cold gas is driven into the inner regions to fuel the central
AGN and trigger a radio outburst. On the other hand, the presence of
fainter larger-scale radio structure surrounding the inner lobes of
Cen A means that in that case at least, the young central radio source
does not represent the first or only radio outburst. Our radio data
show no evidence for previous generations of radio activity in
NGC\,3801.

Radio-loud AGN that are fuelled by the accretion of cold gas from
recent merger events are not self-regulating in the same way as AGN
fuelled by direct accretion of the hot phase of the ISM (e.g. by Bondi
accretion: \citealt{allen}), since the AGN impact on the hot phase
does not affect the accretion rate of cold material. In these systems,
therefore, the long-term effects of AGN energy input may be
particularly extreme.

\section{Conclusions}

We have detected shells of shocked gas surrounding both radio lobes of
NGC\,3801. We infer Mach numbers of $3 - 6$ for the shock advance
speeds, so that our results correspond to the second detection of
strong shocks associated with a low-power radio galaxy. The total
energy stored in the shock-heated shells represents a large fraction
of the total thermal energy of the host galaxy, so that the long-term
effects of the NGC\,3801 outburst are likely to be dramatic. This is
the first case where the total energetic impact of a radio galaxy on
its environment can be directly measured, and suggests that the
energetic contribution of an early supersonic phase is likely to be
important in low-power radio galaxies. The energy stored in the
shocked gas is 25 times the $PdV$ work required to inflate the
radio-lobe cavities, so that during this phase of its evolution, shock
heating is the dominant mechanism of energy transfer from NGC\,3801 to
its environment.

\acknowledgments

We gratefully acknowledge support from the Royal Society (research
fellowship for MJH). This work was partially supported by NASA grant
GO6-7095X. The National Radio Astronomy Observatory is a facility of
the National Science Foundation operated under cooperative agreement
by Associated Universities.

\begin{deluxetable}{lrrrr}
\tablewidth{12cm} 

\tablecaption{Temperature, density and pressure of shells and
surrounding ISM. The ISM values correspond to a distance of 10 arcsec
from the nucleus of NGC\,3801, which is roughly midway along the
shells. For the shell densities and pressures we quote a range
corresponding to the two volume estimates described in
Section~\ref{physical}, and for the ISM density and pressure we quote
in parentheses the range corresponding to the range in abundance and
temperature as described in Section~\ref{ism}. The shell/ISM ratios
for density and pressure are for an ISM abundance of 0.4 times solar.}

\tablehead{Feature&West shell&East shell\\} 
\startdata
$kT_{shell}$ (keV)&$1.0^{+0.4}_{-0.3}$&$0.71^{+0.08}_{-0.04}$\\
$kT_{ISM}$ (keV)&$0.23^{+0.21}_{-0.09}$&$0.23^{+0.21}_{-0.09}$\\
$n_{e,shell}$ (cm$^{-3}$)&$(2.0 - 3.0) \times 10^{-2}$&$(2.0 - 3.0) \times 10^{-2}$\\ 
$n_{e,ISM}$ (cm$^{-3}$)&$4.6 (3.8 - 9.2) \times 10^{-3}$&$4.6 (3.8 - 9.2) \times 10^{-3}$\\ 
$n_{shell}/n_{ism}$&$4.3 - 6.5$&$4.3 - 6.5$\\ 
$P_{shell}$ (Pa)&$(5.9 - 8.9) \times 10^{-12}$&$(4.2 - 6.4) \times 10^{-12}$\\ 
$P_{ISM}$(Pa)&$3.8 (2.9 - 8.9) \times 10^{-13}$&$3.8 (2.9 - 8.9) \times 10^{-13}$\\
$P_{int}$(Pa)&$1.1 \times 10^{-12}$&$1.3 \times 10^{-12}$\\ 
$P_{shell}/P_{ism}$&$15 - 23$&$11-17$\\
$P_{int}/P_{ism}$&$2.9$&$3.4$\\ 
\enddata
\label{properties}
\end{deluxetable}

\begin{figure}
\begin{center}
\plotone{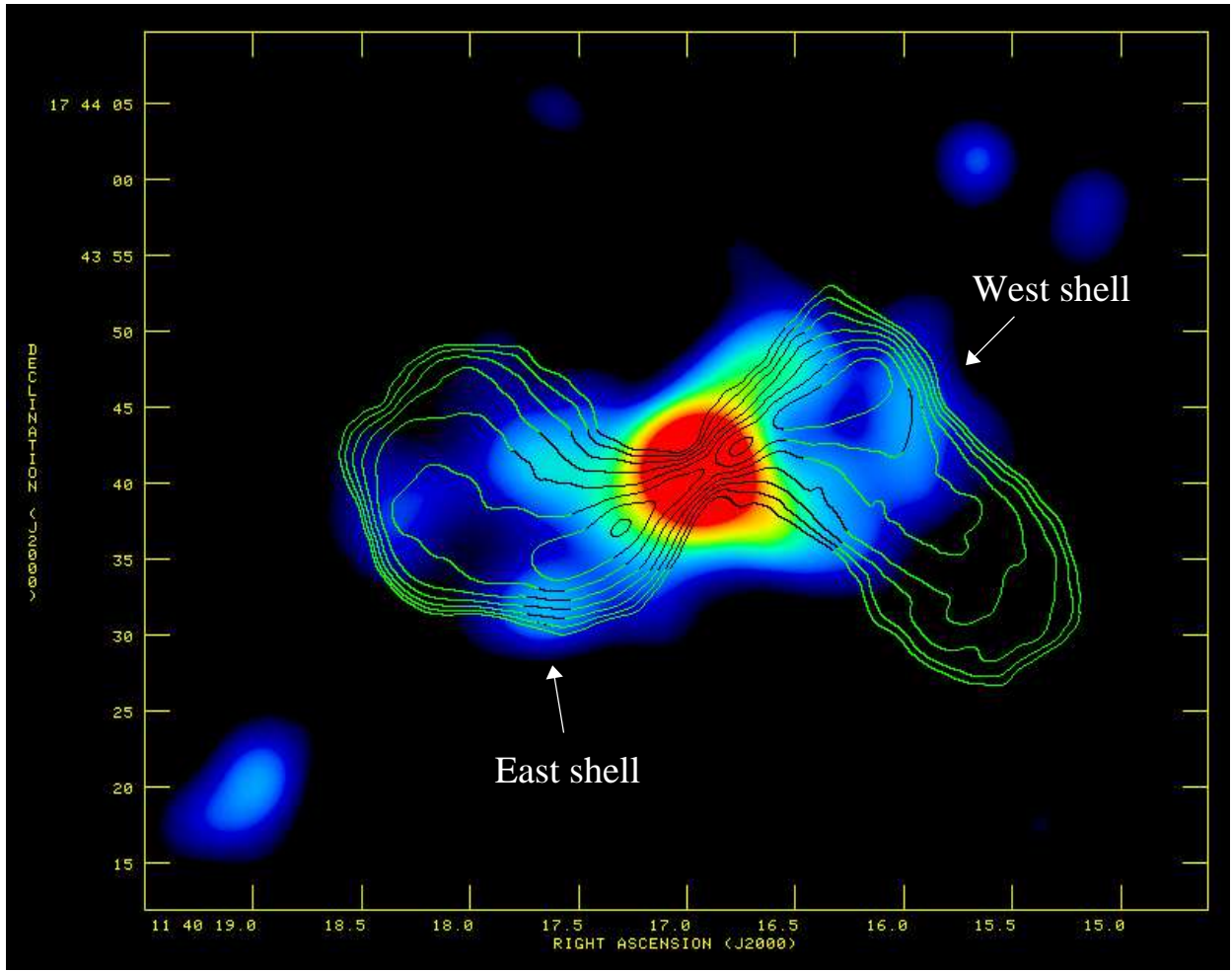}
\caption{Gaussian smoothed ($\sigma = 1.97$ arcsec) 0.5 -- 5 keV image
  of the {\it Chandra} data, with 1.4-GHz radio contours overlaid to
  illustrate the relationship between the X-ray shells and radio
  morphology.}
\label{overlay}
\end{center}
\end{figure}

\end{document}